\documentclass[final]{cvpr}

\usepackage{times}
\usepackage{epsfig}
\usepackage{graphicx}
\usepackage{amsmath}
\usepackage{amssymb}

\usepackage{threeparttable}
\usepackage{booktabs}
\usepackage{stfloats}
\usepackage{bbding}
\usepackage{bm}
\usepackage{multirow}

\usepackage[pagebackref,breaklinks,colorlinks]{hyperref}
\usepackage{color,xcolor}
\usepackage[capitalize]{cleveref}
\colorlet{dark-blue}{blue!70!black}
\hypersetup{
    colorlinks=true,%
    citecolor=dark-blue,%
    filecolor=dark-blue,%
    linkcolor=red,%
    urlcolor=magenta
}
\usepackage{color}
\usepackage{colortbl}
\definecolor{mygray}{gray}{.9}

\def\cvprPaperID{4284} 
\def\confName{CVPR}
\def\confYear{2024}
\graphicspath{{./figs/}{./figs/result/}}

\begin{document}

\title{VoCo: A Simple-yet-Effective Volume Contrastive Learning Framework for 3D Medical Image Analysis}


\author{Linshan Wu\\
\and
Jiaxin Zhuang\\
Hong Kong University of Science and Technology\\
\and
Hao Chen~\thanks{Corresponding author: \href{mailto:jhc@cse.ust.hk}{jhc@cse.ust.hk}}\\
}



\maketitle

\begin{abstract}
   Self-Supervised Learning (SSL) has demonstrated promising results in 3D medical image analysis. However, the lack of high-level semantics in pre-training still heavily hinders the performance of downstream tasks. We observe that 3D medical images contain relatively consistent contextual position information, i.e., consistent geometric relations between different organs, which leads to a potential way for us to learn consistent semantic representations in pre-training. In this paper, we propose a simple-yet-effective \textbf{Vo}lume \textbf{Co}ntrast (\textbf{VoCo}) framework to leverage the contextual position priors for pre-training. Specifically, we first generate a group of base crops from different regions while enforcing feature discrepancy among them, where we employ them as class assignments of different regions. Then, we randomly crop sub-volumes and predict them belonging to which class (located at which region) by contrasting their similarity to different base crops, which can be seen as predicting contextual positions of different sub-volumes. Through this pretext task, VoCo implicitly encodes the contextual position priors into model representations without the guidance of annotations, enabling us to effectively improve the performance of downstream tasks that require high-level semantics. Extensive experimental results on six downstream tasks demonstrate the superior effectiveness of VoCo. Code will be available at \href{https://github.com/Luffy03/VoCo}{https://github.com/Luffy03/VoCo}.

\end{abstract}

\section{Introduction}

\begin{figure}
	\centering
	\includegraphics[width=1\linewidth]{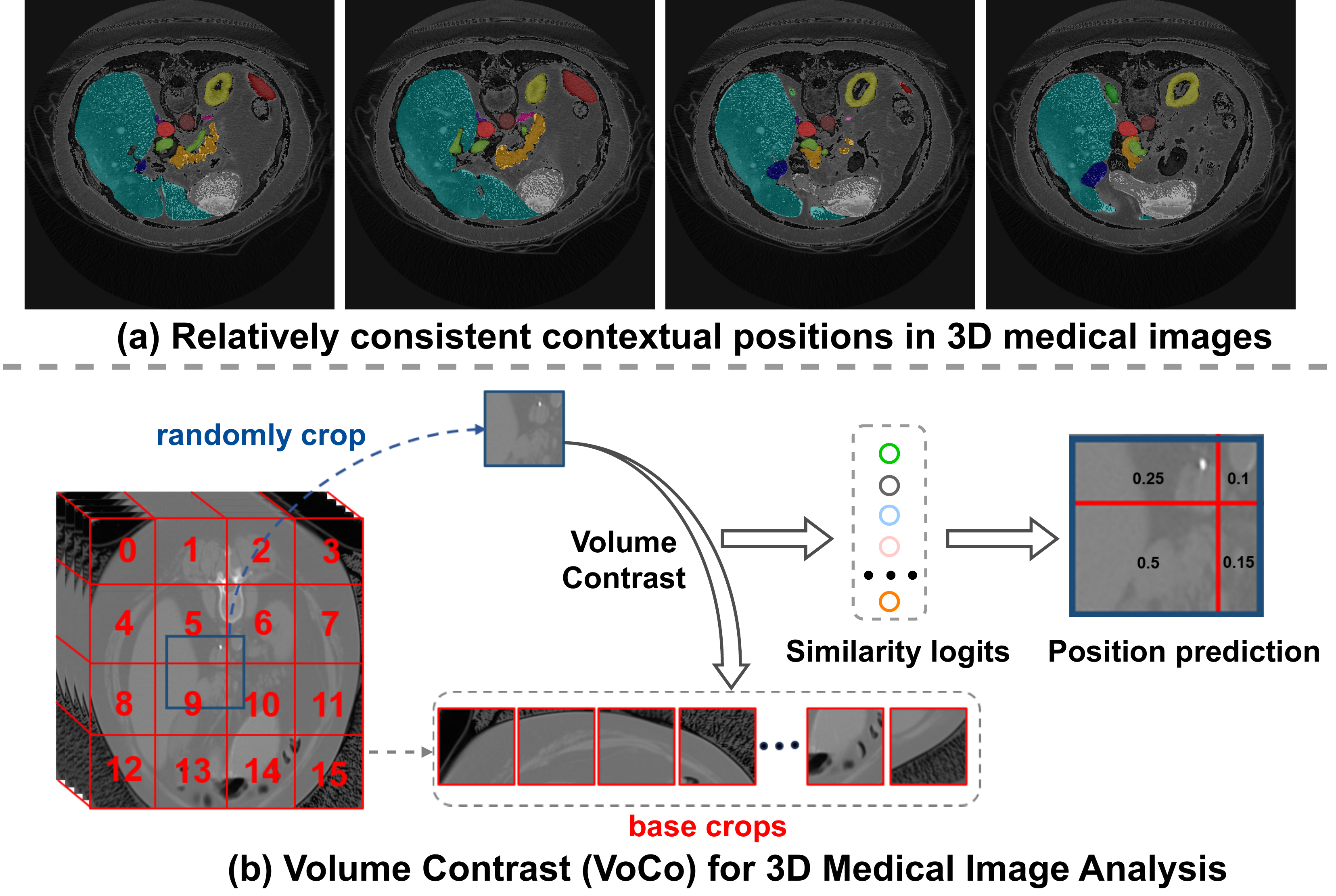}
	\caption{(a) In 3D medical images, the contextual positions, \emph{i.e.}, the geometric relations between different organs are relatively consistent. (b) Aiming to leverage contextual position priors for pre-training, we proposed a Volume Contrast (VoCo) framework for 3D Medical Image Analysis.}
	\label{fig_intro}
\end{figure}


\begin{figure*}
	\centering
	\includegraphics[width=0.8\linewidth]{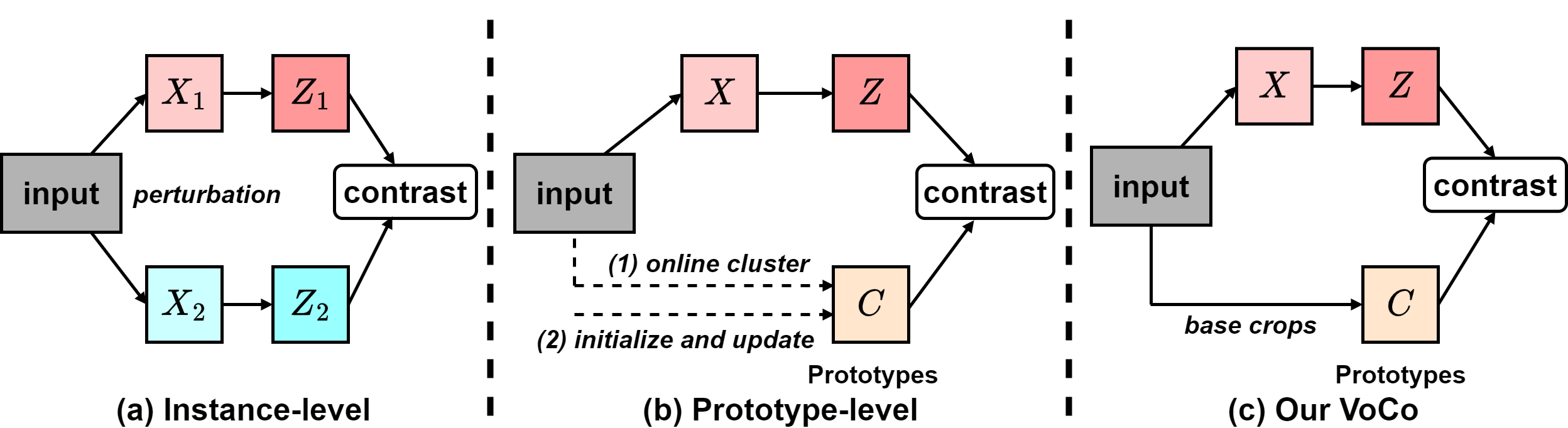}
	\caption{\textbf{Typical contrastive learning frameworks.} (a) Instance-level contrastive learning~\cite{simclr,simsiam,moco,byol,DINO} employs strong data augmentation or model perturbation on input data to acquire different views of instance, then regularizes their consistency. (b) Prototype-level contrastive learning~\cite{deepClu,swav,slot,dinov2,PACO,GPACO} conducts (1) online clustering or (2) randomly initialize then online update process to obtain prototypes as class assignments, then leverage the prototypes to contrast each input image. (c) Our VoCo follows the idea of prototype-level contrastive learning. Specifically, instead of using time-consuming online clustering and updating procedures, we leverage the valuable contextual position priors of 3D medical images and leverage the base crops to generate prototypes (bases).}
	\label{fig_compare}
\end{figure*}

Deep learning has demonstrated outstanding achievements in 3D medical image analysis~\cite{DL1, DL2, DL3, nnunet,word}, yet is heavily hampered by the expensive cost of the required expert annotations~\cite{SSL3D, annotating}. To address this problem, Self-Supervised Learning (SSL) has received significant attention due to its promising ability to learn representations without annotations~\cite{simclr, simsiam, swav, MAE, disco}, which has become an important label-efficient solution in 3D medical image analysis~\cite{PCRLv2,swin,geo,big,Alice,clipdriven}.

Existing methods~\cite{SSL3D, rubik, PCRLv2, MAE3D} are mostly based on information reconstructions to learn augment-invariant representations of 3D medical images, which first employ strong data augmentation to the images and then reconstruct the raw information. Specifically, rotate-and-reconstruct~\cite{SSL3D, swin, rubik, rubik2} proposed to randomly rotate the 3D volumetric images and learn to recover them, which encourages models to learn rotational invariant features. Recent methods~\cite{PCRLv1, PCRLv2,geo,dira,unimiss} further proposed to restore information among different views of the image. PCRL~\cite{PCRLv1, PCRLv2} cropped global and local patches then conducted multi-scale restorations. GVSL~\cite{geo} further explored the geometric similarity between multi-scans by affine augmentation and matching. Mask-reconstruct methods~\cite{MAE3D, GLMAE, swinmm} are also widely used, which are introduced from MAE~\cite{MAE} and aim to learn representations by masking images and reconstructing the missing pixels. Although promising results
have been demonstrated, previous works~\cite{rubik2, geo} have proved that the lack of high-level semantics in pre-training will heavily hinder the performance of downstream tasks. To address this challenge, we argue that stronger high-level semantics should be further involved into 3D medical image pre-training.




To this end, we argue that the contextual position priors of 3D medical images should be further exploited. As shown in Fig.~\ref{fig_intro}(a), we observe that in 3D medical images, different organs (semantic regions) contain relatively consistent contextual positions with relatively consistent anatomic characteristics (shapes). Thus, the consistency of geometric relations between different organs leads to a potential way for us to learn consistent semantic representations for 3D medical images pre-training. In this paper, we propose a pretext task for contextual position predictions, which aims to encode contextual position priors into model representations and enables us to effectively improve the performance of downstream tasks that require high-level semantics.

In this paper, we propose a simple-yet-effective Volume Contrast (VoCo) framework for 3D medical image analysis, as shown in Fig.~\ref{fig_intro}(b). Specifically, we first crop a group of non-overlap volumes from different positions while enforcing feature discrepancy among them. We represent these volumes as a group of bases in the learned high-dimension space, where we employ them as class assignments of different positions. Then, we randomly crop sub-volumes and predict them belonging to which class (located at which position) by contrasting their similarity to different bases, which can be seen as predicting contextual positions of different sub-volumes. In this way, we formulate a contextual position prediction pretext task for 3D medical image SSL. Through learning to predict contextual positions, we implicitly involve the high-level semantic priors into the model representations, which enables us to significantly improve the performance of downstream tasks. Extensive experimental results on six downstream tasks demonstrate that our proposed VoCo clearly outperforms existing state-of-the-art 3D medical image SSL methods. 




\section{Related Works}
\label{sec:relate}

In this section, we first introduce the previous mainstream contrastive learning paradigms. Then, we survey the existing SSL methods for medical image analysis, especially for 3D medical images. Finally, we review the position-related SSL methods for comparisons with our method and highlight the differences. 

\noindent\textbf{Contrastive learning.} Contrastive learning is one of the mainstream paradigms in SSL, which aims to learn consistent representations by contrasting positive and negative pairs of samples without extra annotations~\cite{simclr,swav,moco,simsiam}. According to~\cite{swav}, instance- and prototype-level contrastive learning are two typical types of contrastive learning, as shown in Fig.~\ref{fig_compare}. 


Instance-level contrastive learning~\cite{simclr,simsiam,moco,byol,DINO} transforms input images with different augmentations or model perturbations, aiming to compare the features from each other. Prototype-level contrastive learning~\cite{deepClu,swav,slot,dinov2,PACO,GPACO} proposes to generate prototypes (also called clusters or bases) for contrasting each input image. Specifically, there are two typical ways to generate prototypes. First, Caron \emph{et al.} proposed DeepCluster~\cite{deepClu} to conduct online clustering on the whole dataset to generate prototypes. However, it is very time-consuming to calculate clusters on a large dataset. Thus, some recent works~\cite{swav,slot,PACO,GPACO} propose to randomly initialize a group of prototypes and then update them through back-propagation during training, which has demonstrated promising results. However, there is still no explicit guarantee that these randomly initialized prototypes can be updated well during training.

Our VoCo follows the primary idea of prototype-level constrasive learning. As shown in Fig.~\ref{fig_compare}(c), to address the existing problems mentioned above, instead of randomly initializing and updating prototypes, VoCo leverages the valuable contextual position priors of 3D medical images to generate base crops as prototypes, which also requires no time-consuming clustering on a large dataset.

\noindent\textbf{SSL for medical image analysis.} Due to the high potential in label-efficient learning~\cite{moco,CISC_R,AGMM,wu2024modeling,DBFNet,DCA,liu2023multi}, SSL has also received significant attention in the field of medical image analysis~\cite{PCRLv1,geo,intra,swin,du2023weakly}. Existing methods are mainly based on comparative SSL~\cite{PCRLv2}. Specifically, Zhou \emph{et al.} \cite{C2L} combined Mixup~\cite{mixup} into MoCo~\cite{moco} to learn the diversity of positive and negative samples in InfoNCE~\cite{infonce}. Azizi \emph{et al.} used multi-instance learning to compare multiple views of images from each patient. There are also a number of approaches~\cite{dira,PCRLv1,PCRLv2} that supervising the models via restoring low-level information from raw images.

In 3D medical image analysis, reconstructing raw information is a popular pretext task for learning representations~\cite{SSL3D,swin,PCRLv2}. Existing methods are mainly based on reconstructing information from augmented images. These previous methods first conducted strong data augmentation, \emph{e.g.}, rotate~\cite{swin, rubik, rubik2}, multi-view crops~\cite{PCRLv1, PCRLv2, geo}, and mask~\cite{MAE3D, GLMAE, swinmm}, then supervised the model by reconstructing raw 3D information. Although promising results have been demonstrated, most of these methods still largely ignore the importance of integrating high-level semantics into model representations, which heavily hinders the performance of downstream tasks.

\noindent\textbf{Position-related SSL.} Position-related SSL methods are also explored in a number of previous works~\cite{location,jigsaw,improve,permute,puzzle,contextprediction,position,positionlabel} in the field of natural images. Noroozi \emph{et al.} \cite{puzzle} proposed to predict the order of a set of shuffled patches. Zhai \emph{et al.} \cite{position} and Caron \emph{et al.} \cite{location} proposed to train a ViT~\cite{vit} to predict the locations of each input patch. However, since the geometric relations of different objects are not very consistent in natural images, it is still difficult to effectively learn consistent position representations given visual appearance only (as stated in~\cite{positionlabel}). In addition, previous works~\cite{position,location,positionlabel} mainly trained a linear layer to output the positions directly, which works in a black-box manner.

In this paper, we introduce the pretext task of contextual position prediction into the field of 3D medical images, where the geometric relations between different organs are relatively consistent, which guides us to learn consistent semantic representations in pre-training. Different from the previous methods, in this paper, we introduce a totally different position prediction paradigm. Specifically, instead of using a linear layer to output positions directly, we predict the contextual positions based on volume contrast, which is more intuitive and effective. 

\begin{figure*}
	\centering
	\includegraphics[width=0.8\linewidth]{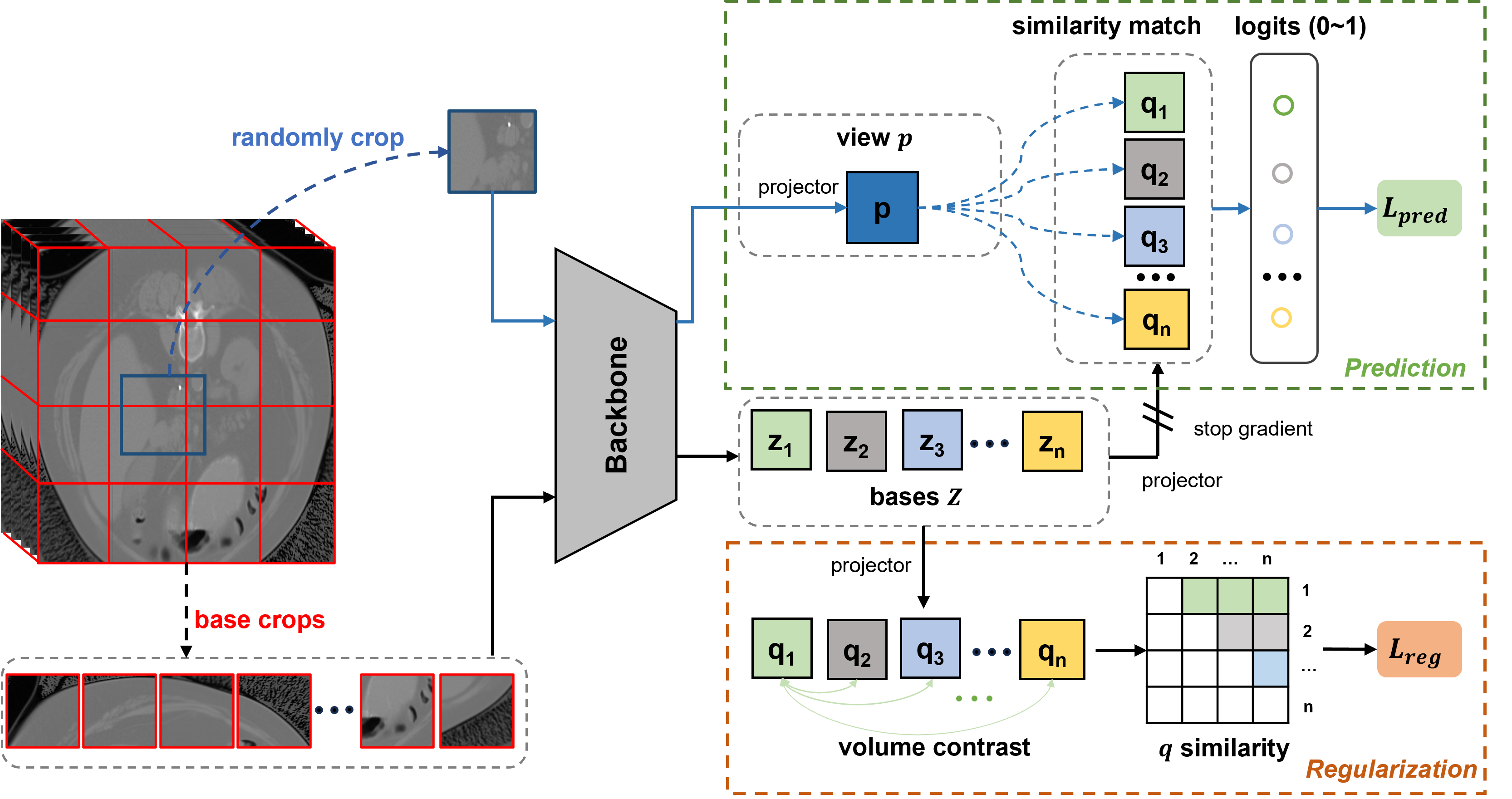}
	\caption{\textbf{The overall framework of VoCo}. VoCo contains a prediction branch and a regularization branch. The prediction branch is responsible for predicting contextual positions between different sub-volumes. The regularization branch is employed to enforce the feature discrepancy between different bases, which aims to learn more discriminative class assignments.}
	\label{fig_frame}
\end{figure*}


\section{Methodology}
\label{sec_method}

In this section, we first introduce the overall framework of our proposed VoCo in Section~\ref{sec3_1}. After that, we present the process of contextual position prediction in Section \ref{sec3_2}. Then, the regularization process via volume contrast in our proposed VoCo framework is described in Section \ref{sec3_3}. 

\subsection{Overall Framework}
\label{sec3_1}

The overall framework of our proposed VoCo is presented in Fig.~\ref{fig_frame}, which contains a contextual position prediction branch and a regularization branch. The prediction branch is used to predict the contextual positions between different cropped volumes. Specifically, given an input volume, we first crop it into non-overlap base volumes, which cover the whole input volume. Then, we randomly crop a volume and transform it into the high-dimension feature space using a typical backbone (CNN~\cite{resnet} or Transformer~\cite{vit}). The goal is to predict the contextual positions between the randomly cropped volumes and base volumes. In this paper, instead of training a linear classifier to predict positions as in previous works~\cite{position,jigsaw,location,positionlabel}, we propose to establish this goal by volume contrast. We develop a loss function $L_{pred}$ to supervise the final predictions. In addition, we further use a loss function $L_{reg}$ to regularize the feature discrepancy from different bases by enlarging their distance, aiming to learn more discriminative class assignments. The details are presented in Section~\ref{sec3_2} and \ref{sec3_3}.

\subsection{Contextual Position Prediction}
\label{sec3_2}

\noindent\textbf{Base and random crops}. Given an input volume, we first crop it into $n$ non-overlap base volumes, which cover the whole input volume. We then employ the extracted features $z$ as class assignments (we call them bases), which present the prototype-level features from different positions. Then, following previous SSL works~\cite{simclr,simsiam,moco}, a projector with linear layers is used to project $z$ into latent features $q$. Then, we randomly crop a volume and transform it into high-dimension feature space as $p$. The backbone and projector are also used to project the features from the randomly cropped volumes.

\noindent\textbf{Volume contrast for contextual position prediction}. With features extracted from the backbone and projector, following previous SSL works~\cite{simclr,simsiam,moco}, we first conduct 3D adaptive average pooling to resize them to one dimension, \emph{i.e.}, $p{\in}{\mathbb{R}}^{1{\times}{C}}$ and $q{\in}{\mathbb{R}}^{1{\times}{C}}$, where $C$ is the number of channels. Specifically, we empirically set $C$ to $2048$ as in~\cite{simclr,simsiam,moco}. 

Then, we calculate the similarity logits $l$ between $p$ and ${q}_{i}$. Specifically, we use cosine similarity to compute $l$ as follows:
\begin{equation}\label{logit}
	l_{i}=CosSim(p, {q}_{i})=\frac{p{\cdot}{{q}_{i}}}{\parallel p\parallel \parallel {q}_{i} \parallel}, i{\in}n
\end{equation}
where ${q}_{i}$ is the projected feature of each base crop. $l_{i}$ denotes the similarity between $p$ in ${q}_{i}$, which ranges from $0$ to $1$. It is worth noting that, we stop the gradients of $q$ when computing Eq.~\ref{logit}, which aims to avoid feature collapse~\cite{simclr,simsiam,swav}.

Intuitively, higher $l_{i}$ represents that $p$ has higher probabilities to share overlap regions with ${q}_{i}$. In this way, we can explicitly associate the similarity value with the position information, \emph{i.e.}, $p$ with higher ${l}_{i}$ is more likely to be located in the region of the $i_{th}$ base. Thus, instead of training a black-box linear layer, we predict the contextual positions by volume contrast, which is more intuitive and effective.

\begin{figure}
	\centering
	\includegraphics[width=0.8\linewidth]{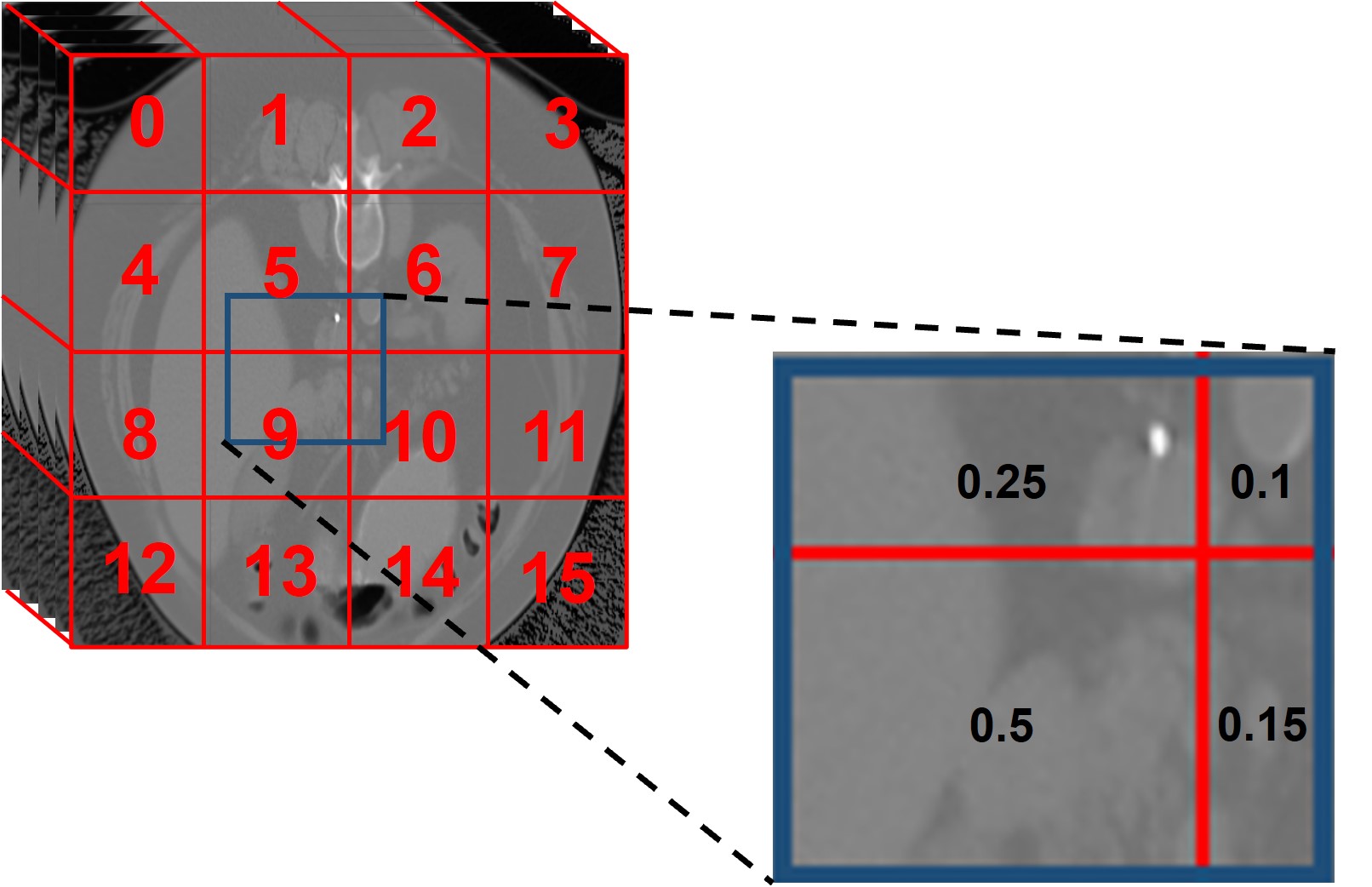}
	\caption{\textbf{The process of generating position labels}. We calculate the proportions of overlap area as position labels $y$, \emph{e.g.}, the randomly cropped volume in the figure is assigned to class 5, 6, 9, and 10 with probabilities of 0.25, 0.1, 0.5, and 0.15, respectively.}
	\label{fig_label}
\end{figure}

\noindent\textbf{Position labels generation}. The process of generating position labels is shown in Fig.~\ref{fig_label}. As shown in Fig.~\ref{fig_label}, when we generate $n=4{\times}4$ base crops, there will be $n$ class assignments. Then we calculate the overlap area between a randomly cropped volume and $n$ base crops. The proportions of the overlap area are then assigned as position labels $y$, which also range from $0$ to $1$. Thus, we can easily supervise the model by calculating the distance between the prediction logits $l$ and position labels $y$. The setting of the number $n$ of base crops will be discussed in Section~\ref{sec_ablation}.

\noindent\textbf{Loss function for contextual position prediction}. The formulation of prediction loss function $L_{pred}$ is based on entropy. Specifically, we first calculate the distance $d$ between prediction logits $l$ and position labels $y$:
\begin{equation}\label{dist}
	d_{i}=|{y}_{i}-{l}_{i}|, i{\in}n,
\end{equation}
where $|.|$ denotes the absolute value. Then, $L_{pred}$ is formulated as follows:
\begin{equation}\label{L_pred}
	L_{pred}=-\frac{1}{n}\sum_{i{\in}n}^{n}log(1-d_{i}).
\end{equation}

It is worth noting that VoCo predicts contextual positions of a volume (high similarities with all its contextual overlapped volumes), thus don't need one-to-one correspondence: \emph{e.g.}, in Fig.~\ref{fig_label}, high-value ${l}_{i}$ pertain to ${y}_{i}{>}0{,}i{=}5,6,9,10$ simultaneously. Then we calculate the distance between $l_{i}$ and $y_{i}$ (Eq.~\ref{dist}).

\subsection{Volume Contrast for Regularization}
\label{sec3_3}

We aim to learn more discriminative class assignments (bases) for volume contrast. Since intuitively, different sub-volumes tend to contain different organs (semantic discrepancy). Thus, we aim to enlarge the high-dimension feature discrepancy between different bases. To this end, we design a regularization loss $L_{reg}$ to enlarge the feature discrepancy between different bases $z$. 

First, given projected bases $q$, we also calculate the cosine similarity $s_{ij}$ between different $q_{i}$ and $q_{j}$ as follows:
\begin{equation}\label{cos}
	s_{ij}=CosSim({q}_{i}, {q}_{j})=\frac{{q}_{i}{\cdot}{{q}_{j}}}{\parallel {q}_{i}\parallel \parallel {q}_{j} \parallel}, i,j{\in}n, i{\neq}j,
\end{equation}
where we aim to regularize $s_{ij}$ to $0$, enforcing feature discrepancy between different bases. Thus, the loss function $L_{reg}$ is formulated as:
\begin{equation}\label{L_reg}
	L_{reg}=\frac{2}{n(n-1)}\sum_{i,j{\in}n,i{\neq}j}^{n}|s_{ij}|, 
\end{equation}
where $|.|$ denotes the absolute value. With loss $L_{reg}$, we aim to optimize $q$ as linearly independent bases:
\begin{equation}\label{opti}
	{q}_{i} {\perp} {q}_{j}, i{\neq}j, i,j{\in}n.
\end{equation}

With the regularization loss function $L_{reg}$, we aim to learn a group of linearly independent bases to represent all directions of high-dimension features~\cite{swav}. In this way, we can learn a group of more discriminative class assignments to supervise the final position predictions.

\noindent\textbf{Overall loss function}. Thus, the total loss function $L$ is the combination of $L_{reg}$ and $L_{pred}$:
\begin{equation}\label{L_total}
	L = L_{pred} + {\lambda}L_{reg},
\end{equation}
where $\lambda$ is used to balance the relative contributions of these two loss terms and set to $1.0$ in experiments empirically since we consider their importance equally. The ablation studies of $\lambda$ are provided in the supplementary materials.

\section{Experiments}
\label{sec4}
In this section, we first describe the datasets used in the pre-training and downstream tasks. Then, we briefly introduce the implementation details of VoCo. Finally, we report detailed experiment results of our proposed VoCo compared with other state-of-the-art SSL methods in 3D medical images. More details are in the supplementary materials.

\subsection{Datasets}\label{sec4_1}

\noindent\textbf{Pre-training datasets}. Aiming to conduct fair comparisons with the previous works~\cite{swin,swinmm,PCRLv1,PCRLv2,MAE3D,GLMAE}, we also carry out pre-training experiments on the same three public datasets, \emph{i.e.}, BTCV~\cite{btcv}, TCIA Covid19~\cite{tcia}, and LUNA~\cite{luna} datasets, including about totally 1.6k CT scans for pre-training. It is worth noting that, aiming to conduct fair comparisons with previous works~\cite{GLMAE,MAE3D}, we only use BTCV~\cite{btcv} and TCIA Covid-19~\cite{tcia} for pre-training in the downstream experiments of BTCV~\cite{btcv}. For the other downstream tasks, we use all three datasets for pre-training. Details are provided in the supplementary materials.

\noindent\textbf{Downstream datasets}. To evaluate the effectiveness of our VoCo, we conduct downstream experiments on six public datasets, \emph{i.e.}, BTCV~\cite{btcv}, LiTs~\cite{lits}, MSD Spleen~\cite{MSD}, MM-WHS~\cite{mmwhs}, BraTS 21~\cite{brats}, and CC-CCII~\cite{CC-CCII}, including segmentation and classification tasks. The first five datasets are developed for segmentation, while CC-CCII~\cite{CC-CCII} is for COVID-19 classification. Note that only BTCV~\cite{btcv} is used in pre-training, the other datasets are unseen in pre-training. In addition, to evaluate the cross-modality generalization ability, we transfer the model pre-trained on the CT dataset to the MRI dataset BraTS 21~\cite{brats}. We adopt consistent settings as previous works~\cite{MAE3D,GLMAE,swinunetr,swinmm,geo}. We also evaluate the performance on 2D medical dataset~\cite{chestx-ray}. Details are provided in the supplementary materials.

\begin{table*}
	\setlength{\abovecaptionskip}{0.pt}
	\setlength{\belowcaptionskip}{-0.em}
	\centering
	\footnotesize
\begin{threeparttable}
	\begin{tabular}{ccccccccccccccc}
		\toprule[1.2pt]
		\multirow{2}{*}{\textbf{Method}} & \multicolumn{14}{c}{\textbf{Dice Score(\%)}}\\
		\cline{2-14}
        &Spl &RKid &LKid &Gall &Eso &Liv &Sto &Aor &IVC &Veins &Pan &RAG &LAG &\textbf{AVG}\\
        \hline
        \textbf{\emph{From Scratch}}\\
		UNETR~\cite{unetr} &93.02 &94.13 &94.12 &66.99 &70.87 &96.11 &77.27 &89.22 &82.10 &70.16 &76.65 &65.32 &59.21 &79.82\\
        Swin-UNETR$\dagger$~\cite{swinunetr} &94.06 &93.54 &93.80 &65.51 &74.60 &97.09 &75.94 &91.80 &82.36 &73.63 &75.19 &68.00 &61.11 &80.53\\
        \hline
        \textbf{\emph{With General SSL}}\\
        MAE3D~\cite{MAE,MAE3D} &93.98 &94.37 &94.18 &69.86 &74.65 &96.66 &80.40 &90.30 &83.13 &72.65 &77.11 &67.34 &60.54 &81.33\\
        SimCLR~\cite{simclr} &92.79 &93.04 &91.41 &49.65 &50.99 &98.49 &77.92 &85.56 &80.58 &64.37 &67.16 &59.04 &48.99 &73.85\\
        SimMIM~\cite{simmim} &95.56 &95.82 &94.14 &52.06 &53.52 &\textbf{98.98} &80.25 &88.11 &82.98 &66.49 &69.16 &60.88 &50.45 &76.03\\
        MoCo v3$\dagger$~\cite{moco,mocov3} &91.96 &92.85 &92.42 &68.25 &72.77 &94.91 &78.82 &88.21 &81.59 &71.15 &75.76 &66.48 &58.81 &79.54\\
        Jigsaw$\dagger$~\cite{jigsaw} &94.62 &93.41 &93.55 &75.63 &73.21 &95.71 &80.80 &89.41 &84.78 &71.02 &79.57 &65.68 &60.22 &81.35\\
        PositionLabel$\dagger$~\cite{positionlabel} &94.35 &93.15 &93.21 &75.39 &73.24 &95.76 &80.69 &88.80 &84.04 &71.18 &79.02 &65.11 &60.07 &81.09\\
        \hline
        \textbf{\emph{With Medical SSL}}\\
        MG~\cite{Modelgen} &91.99 &93.52 &91.81 &65.11 &76.14 &95.98 &\textbf{86.88} &89.29 &83.59 &71.79 &81.62 &67.97 &63.18 &81.45\\
        ROT~\cite{SSL3D} &91.75 &93.13 &91.62 &65.09 &\textbf{76.55} &94.21 &86.16 &89.74 &83.08 &71.13 &81.55 &67.90 &63.72 &81.20\\
        Vicregl~\cite{vicregl} &90.32 &94.15 &91.30 &65.12 &75.41 &94.76 &86.00 &89.13 &82.54 &71.26 &81.01 &67.66 &63.08 &80.89\\
		Rubik++$\dagger$~\cite{rubik2} &\textbf{96.21} &90.41 &89.33 &75.22 &72.64 &97.44 &79.25 &89.65 &83.76 &74.74 &78.35 &67.14 &61.97 &81.38\\
        PCRLv1$\dagger$~\cite{PCRLv1} &95.73 &89.66 &88.53 &75.41 &72.33 &96.20 &78.99 &89.11 &83.06 &74.47 &77.88 &67.02 &61.85 &80.78\\
        PCRLv2$\dagger$~\cite{PCRLv2} &95.50 &91.43 &89.52 &\textbf{76.15} &73.54 &97.28 &79.64 &90.16 &84.17 &75.20 &78.71 &68.74 &62.93 &81.74\\
		Swin-UNETR~\cite{swinunetr,swin} &95.25 &93.16 &92.97 &63.62 &73.96 &96.21 &79.32 &89.98 &83.19 &76.11 &\textbf{82.25} &68.99 &65.11 &81.54\\
        SwinMM~\cite{swinmm} &94.33 &94.18 &94.16 &72.97 &74.75 &96.37 &83.23 &89.56 &82.91 &70.65 &75.52 &69.17 &62.90 &81.81\\
        GL-MAE~\cite{GLMAE} &94.54 &94.39 &94.37 &73.19 &74.93 &96.51 &83.49 &89.74 &83.11 &70.80 &75.71 &69.39 &63.12 &82.01\\
        GVSL$\dagger$~\cite{geo} &95.27 &91.22 &92.25 &72.69 &73.56 &96.44 &82.40 &88.90 &84.22 &70.84 &76.42 &67.48 &63.25 &81.87\\
        \rowcolor{mygray}
        \textbf{VoCo} &95.73 &\textbf{96.53} &\textbf{94.48} &76.02 &75.60 &97.41 &78.43 &\textbf{91.21} &\textbf{86.12} &\textbf{78.19} &80.88  &\textbf{71.47} &\textbf{67.88} &\textbf{83.85}\\
		\toprule[1.2pt]
	\end{tabular}
    \end{threeparttable}       
	\caption{Experimental results on BTCV~\cite{btcv}. The best results are \textbf{bolded}. `From Scratch' denotes the supervised baseline without self-supervised pre-training. $\dagger$ denotes we re-implement the approach. Most results are drawn from~\cite{MAE3D,dive,GLMAE} or their own papers.}
\label{table_BTCV}
\end{table*}%

\begin{table}
	\setlength{\abovecaptionskip}{0.pt}
	\setlength{\belowcaptionskip}{-0.em}
	\centering
	\footnotesize
\begin{threeparttable}
	\begin{tabular}{ccc}
		\toprule[1.2pt]
		\textbf{Method} &\textbf{Network} &\textbf{Dice Score(\%)}\\
		\hline
        \textbf{\emph{From Scratch}}\\
        3D UNet~\cite{UNET} &- &90.70\\
        UNETR$\dagger$~\cite{unetr} &- &93.25\\
        Swin-UNETR$\dagger$~\cite{swinunetr} &- &93.42\\
        \hline
        \textbf{\emph{With General SSL}}\\
        Jigsaw~\cite{jigsaw} &3D UNet &94.36\\
        MAE3D~\cite{MAE,MAE3D} &UNETR &94.02\\
        MoCo v3$\dagger$~\cite{moco,mocov3} &UNETR &93.86\\
        Jigsaw$\dagger$~\cite{jigsaw} &Swin-UNETR &95.24\\
        PositionLabel$\dagger$~\cite{positionlabel} &Swin-UNETR &94.13\\
        \hline
        \textbf{\emph{With Medical SSL}}\\
        MG~\cite{Modelgen} &3D UNet &91.30\\
        TransVW~\cite{TransVW} &3D UNet &91.42\\
        ROT~\cite{SSL3D} &3D UNet &94.49\\
        PCRLv1~\cite{PCRLv1} &3D UNet &93.87\\
		PCRLv2~\cite{PCRLv2} &3D UNet &94.50\\
		Rubik~\cite{rubik} &3D UNet &94.93\\
		Rubik++~\cite{rubik2} &3D UNet &95.46\\
        Rubik++$\dagger$~\cite{rubik2} &Swin-UNETR &95.72\\
        Swin-UNETR~\cite{swin,swinunetr} &Swin-UNETR &95.33\\
        SwinMM~\cite{swinmm} &Swin-UNETR &95.52\\
        \rowcolor{mygray}
        \textbf{VoCo} &3D UNet &\textbf{96.03}\\
		\rowcolor{mygray}
        \textbf{VoCo} &Swin-UNETR &\textbf{96.52}\\
        \toprule[1.2pt]
	\end{tabular}
    \end{threeparttable}        
	\caption{Experimental results on LiTs~\cite{lits}. We report the Dice Scores of liver segmentation. $\dagger$ denotes we re-implement the approach. Most results are drawn from~\cite{dive,PCRLv2}.}
\label{table_lits}
\end{table}%

\begin{table}
	\setlength{\abovecaptionskip}{0.pt}
	\setlength{\belowcaptionskip}{-0.em}
	\centering
	\footnotesize
\begin{threeparttable}
	\begin{tabular}{ccc}
		\toprule[1.2pt]
		\textbf{Method} &\textbf{MSD Spleen} &\textbf{MM-WHS}\\
		\hline
        \textbf{\emph{From Scratch}}\\
        3D UNet~\cite{UNET} &93.71 &83.09\\
        UNETR$\dagger$~\cite{unetr} &94.20 &85.85\\
        Swin-UNETR$\dagger$~\cite{swinunetr} &94.63 &86.11\\
        \hline
        \textbf{\emph{With General SSL}}\\
        MAE3D~\cite{MAE,MAE3D} &95.20 &86.03\\
        MoCo v3$\dagger$~\cite{moco,mocov3} &94.32 &84.16\\
        Jiasaw$\dagger$~\cite{jigsaw} &94.29 &85.88\\
        PositionLabel$\dagger$~\cite{positionlabel} &94.16 &85.52\\
        \hline
        \textbf{\emph{With Medical SSL}}\\
        MG~\cite{Modelgen} &94.40 &86.36\\
        VicRegl~\cite{vicregl} &94.12 &84.72\\
        UniMiss~\cite{unimiss} &95.09 &84.68\\
        PCRLv1$\dagger$~\cite{PCRLv1} &94.32 &86.58\\
        PCRLv2$\dagger$~\cite{PCRLv2} &94.94 &86.82\\
        Rubik++$\dagger$~\cite{rubik2} &95.11 &87.02\\
        Swin-UNETR~\cite{swin,swinunetr} &95.02 &87.06\\
        SwinMM~\cite{swinmm} &95.34 &86.98\\
        JSSL~\cite{JSSL} &94.92 &84.89\\
        GVSL~\cite{geo} &95.47 &88.27\\
        \rowcolor{mygray}
        \textbf{VoCo} &\textbf{96.34} &\textbf{90.54}\\
        \toprule[1.2pt]
	\end{tabular}
    \end{threeparttable}        
	\caption{Experimental results on MSD Spleen~\cite{MSD} and MM-WHS~\cite{mmwhs}. We report the Dice Scores of segmentation prediction. $\dagger$ denotes we re-implement the approach.}
\label{table_spleen_mm}
\end{table}%

\begin{table}
	\setlength{\abovecaptionskip}{0.pt}
	\setlength{\belowcaptionskip}{-0.em}
	\centering
	\footnotesize
\begin{threeparttable}
	\begin{tabular}{cccccc}
		\toprule[1.2pt]
		\multirow{2}{*}{\textbf{Method}} &\multirow{2}{*}{\textbf{Net.}} &\multicolumn{4}{c}{\textbf{Dice Score(\%)}}\\
        \cline{3-6}
        & &TC &WT &ET &\textbf{AVG}\\
		\hline
        \textbf{\emph{From Scratch}}\\
        UNETR~\cite{unetr} &- &81.62 &87.81 &57.34 &75.58\\
        Swin-UNETR~\cite{swinunetr} &- &81.28 &88.67 &57.73 &75.89\\
        \hline
        \textbf{\emph{With General SSL}}\\
        MAE3D~\cite{MAE,MAE3D} &UNETR &82.34 &90.35 &59.18 &77.29\\
        SimMIM~\cite{simmim} &UNETR &84.06 &90.43 &59.07 &77.85\\
        SimCLR~\cite{simclr} &UNETR &83.13 &89.44 &58.42 &76.99\\
        MoCo v3$\dagger$~\cite{moco,mocov3} &UNETR &82.60 &88.89 &57.69 &76.39\\
        Jigsaw$\dagger$~\cite{jigsaw} &Sw-UNE. &81.62 &89.45 &59.10 &76.72\\
        PositionLabel$\dagger$~\cite{positionlabel} &Sw-UNE. &81.35 &89.62 &58.73 &76.64\\
        \hline
        \textbf{\emph{With Medical SSL}}\\
        PCRLv1$\dagger$\cite{PCRLv1} &Sw-UNE. &81.96 &88.83 &57.58 &76.12\\
        PCRLv2$\dagger$\cite{PCRLv2} &Sw-UNE. &82.13 &90.06 &57.70 &76.63\\
        Rubik++$\dagger$\cite{rubik2} &Sw-UNE. &84.32 &90.23 &58.01 &77.51\\
        Swin-UNETR~\cite{swin,swinunetr} &Sw-UNE. &82.51 &89.08 &58.15 &76.58\\
        SwinMM~\cite{swinmm} &Sw-UNE. &83.48 &\textbf{90.47} &58.72 &77.56\\
		\rowcolor{mygray}
        \textbf{VoCo} &Sw-UNE. &\textbf{85.27} &90.45 &\textbf{59.87} &\textbf{78.53}\\
        \toprule[1.2pt]
	\end{tabular}
    \end{threeparttable}        
	\caption{Experimental results on BraTS 21~\cite{brats}. WT, TC, and ET denote the whole tumor, tumor core, and enhancing tumor, respectively. $\dagger$ denotes we re-implement the approach. }
\label{table_brats}
\end{table}%

\begin{table}
	\setlength{\abovecaptionskip}{0.pt}
	\setlength{\belowcaptionskip}{-0.em}
	\centering
	\footnotesize
\begin{threeparttable}
	\begin{tabular}{ccc}
		\toprule[1.2pt]
		\textbf{Method} &\textbf{Network} &\textbf{Accuracy(\%)}\\
		\hline
        \textbf{\emph{From Scratch}}\\
        UNETR~\cite{unetr} &- &88.92\\
        Swin-UNETR~\cite{swinunetr} &- &88.04\\
        \hline
        \textbf{\emph{With General SSL}}\\
        MAE3D~\cite{MAE, MAE3D} &UNETR &89.47\\
        MoCo v3~\cite{moco,mocov3} &UNETR &84.95\\
        Jiasaw~\cite{jigsaw} &Swin-UNETR &86.88\\
        PositionLabel~\cite{positionlabel} &Swin-UNETR &87.54\\
        \hline
        \textbf{\emph{With Medical SSL}}\\
        PCRLv1~\cite{PCRLv1} &Swin-UNETR &88.72\\
        PCRLv2~\cite{PCRLv2} &Swin-UNETR &89.15\\
        Rubik++~\cite{rubik2} &Swin-UNETR &89.23\\
        Swin-UNETR~\cite{swin,swinunetr} &Swin-UNETR &89.45\\
        SwinMM~\cite{swinmm} &Swin-UNETR &89.61\\
        \rowcolor{mygray}
        \textbf{VoCo} &Swin-UNETR &\textbf{90.83}\\
        \toprule[1.2pt]
	\end{tabular}
    \end{threeparttable}        
	\caption{Experimental results of CC-CCII~\cite{CC-CCII} classification.}
\label{table_cc}
\end{table}%

\subsection{Implementation details}\label{sec4_2} 

Following previous works~\cite{swin,swinmm}, we use SwinUNETR~\cite{swinunetr} for both pre-training and downstream tasks. We use AdamW~\cite{adamw} optimizer and cosine learning rate scheduler for all experiments. We set 100K training steps in the pre-training process and applied a slicing window inference for fair comparisons with previous works~\cite{swin,swinmm,MAE3D,GLMAE}. Aiming to evaluate the pure effectiveness, we do not use foundation models or post-processing~\cite{Alice,clipdriven}. Details are provided in the supplementary materials.


\noindent\textbf{Comparison methods}. We compare our VoCo with both General and Medical SSL methods. First, we compare with the typical SSL methods MAE~\cite{MAE, MAE3D} and MoCo v3~\cite{moco, mocov3}, since they represent the two mainstream SSL paradigms, \emph{i.e.}, mask-autoencoder and constrastive learning. Since it is not practical to set a large batch size for 3D medical images due to computation cost, for fair comparisons, we adopt consistent settings with other methods in MAE~\cite{MAE, MAE3D} and MoCo v3~\cite{moco, mocov3}. We also report the results of SimCLR~\cite{simclr} and SimMIM~\cite{simmim} according to~\cite{MAE3D}. We further evaluate the performance of Jiasaw~\cite{jigsaw} and PositionLabel~\cite{positionlabel}, since they are related to our position-aware methods. Most existing state-of-the-art medical SSL methods are compared in our experiments.

\subsection{Experiments on downstream tasks}\label{sec4_3}

\noindent\textbf{Outperform existing methods on the BTCV dataset}. We first conduct experiments on BTCV~\cite{btcv}, as shown in Table~\ref{table_BTCV}. Specifically, among the comparison methods, MAE3D~\cite{MAE,MAE3D}, SimCLR~\cite{simclr}, SimMIM~\cite{simmim}, MoCo v3~\cite{moco,mocov3}, and GL-MAE~\cite{GLMAE} use UNETR~\cite{unetr}. Other methods including our VoCo adopt Swin-UNETR~\cite{swinunetr} as the default settings of the previous work~\cite{swin}. 

\textbf{Remark}. It can be seen in Table~\ref{table_BTCV} that the general SSL methods perform worse than most medical SSL methods. Specifically, MoCo v3~\cite{moco,mocov3} can only achieve 79.54\% Dice Score. Since MoCo v3~\cite{moco,mocov3} heavily relies on a large batch size to acquire adequate negative samples, which is not practical in 3D medical images due to the huge computation burden. In addition, the negative relation between different images used in MoCo~\cite{moco,mocov3} is not appropriate in medical images. MAE~\cite{MAE,MAE3D}, SimCLR~\cite{simclr}, and SimMIM~\cite{simmim} (results from~\cite{MAE3D}) also gain limited performance. Our VoCo also outperforms the position-based methods Jigsaw~\cite{jigsaw} and PositionLabel~\cite{positionlabel} by a clear margin. Thus, we conclude that general SSL methods are not very suitable for 3D medical images. It is crucial to consider the characteristics of medical images in medical SSL. 

The scratch Swin-UNETR~\cite{swinunetr} only achieves 80.53\% Dice Score. With VoCo pre-training, we gain 3.32\% improvements with 83.85\% Dice Score, which also outperforms existing methods by a clear margin. Among the compared methods, GL-MAE~\cite{GLMAE} achieves the highest Dice Score (82.01\%). Our VoCo surpasses it by 1.84\% Dice Score, which is a clear improvement in this dataset. 

\noindent\textbf{Promising performance on Unseen datasets}. We further conduct experiments on unseen datasets in pre-training, \emph{i.e.}, LiTs~\cite{lits}, MSD Spleen~\cite{MSD}, and MM-WHS~\cite{mmwhs}. The results on LiTs~\cite{lits} are shown in Table~\ref{table_lits}. We report the results of compared methods according to~\cite{PCRLv1,PCRLv2,dive}. Since the scratch Swin-UNETR~\cite{swinunetr} can obtain a higher Dice Score (93.42\%), we further pre-train a 3D UNet~\cite{UNET} based on VoCo, aiming to conduct fair comparisons. It can be seen that with VoCo pre-training, Swin-UNETR~\cite{swinunetr} gains 3.10\% improvements and achieves 96.52\% Dice Score. With 3D UNet~\cite{UNET} as the backbone, VoCo also achieves 96.03\% Dice Score, proving the effectiveness of VoCo with different network architectures. 

\begin{figure}
	\centering
	\includegraphics[width=1\linewidth]{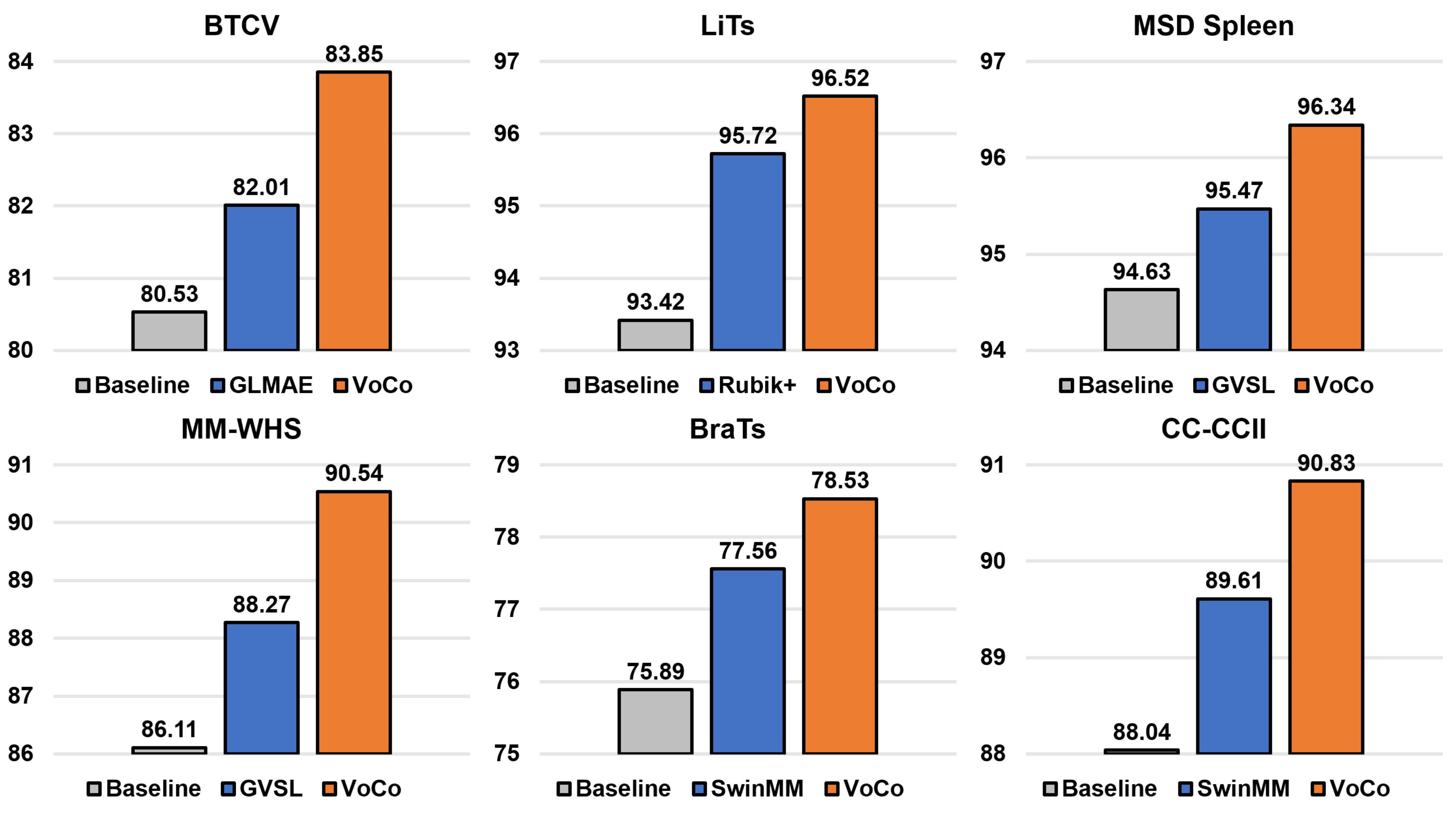}
	\caption{Overall comparisons with state-of-the-art methods on six different datasets.}
	\label{fig_over}
\end{figure}

\begin{figure*}
	\centering
	\includegraphics[width=0.65\linewidth]{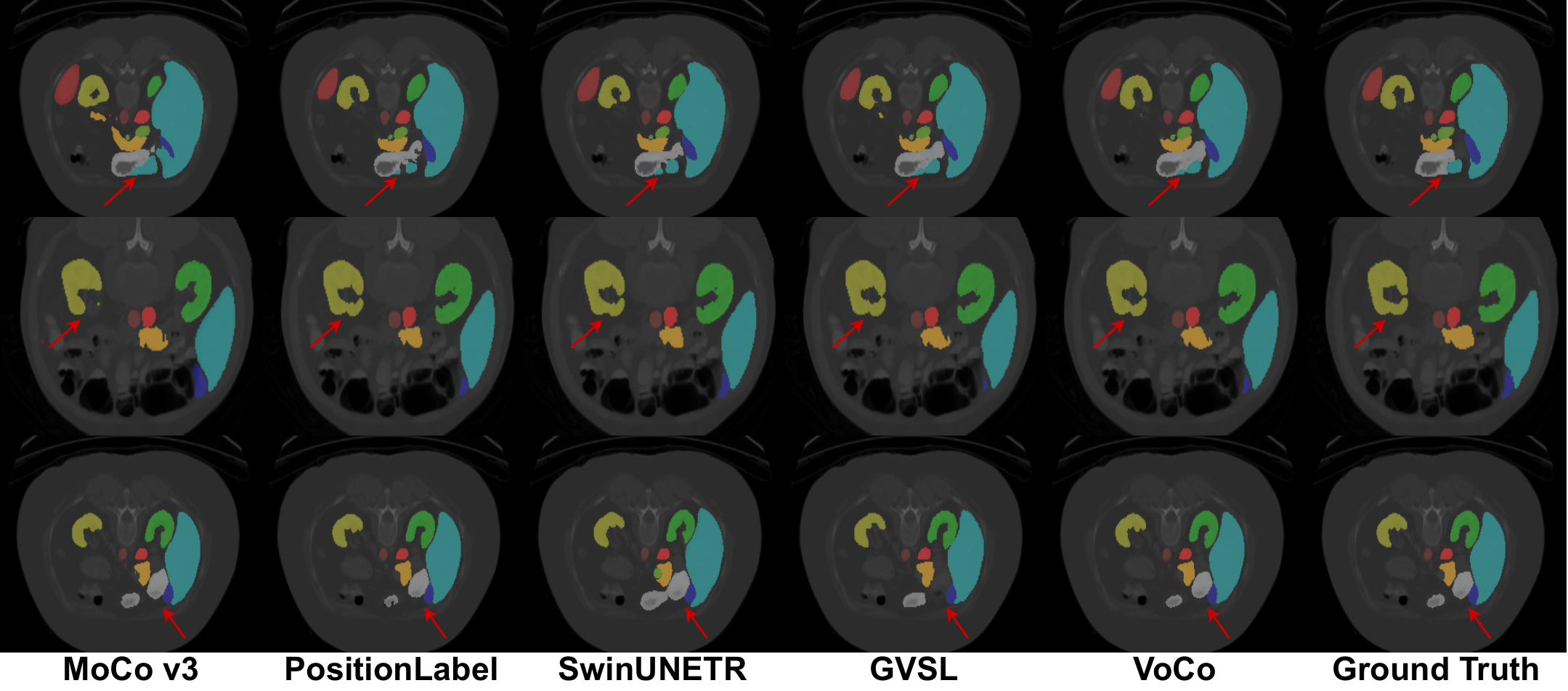}
	\caption{Qualitative visualization of segmentation results for the BTCV~\cite{btcv} dataset. We compare VoCo with MoCo v3~\cite{mocov3}, PositionLabel~\cite{positionlabel}, SwinUNETR~\cite{swin, swinunetr}, and GVSL~\cite{geo}.}
	\label{fig_btcv_vis}
\end{figure*}



The results on MSD Spleen~\cite{MSD} and MM-WHS~\cite{mmwhs} datasets are shown in Table~\ref{table_spleen_mm}. In previous methods, GVSL~\cite{geo} achieves the best performance with 95.47\% and 88.27\% Dice Score, while our VoCo surpasses all previous methods with 96.34\% and 90.54\% Dice Score on MSD Spleen~\cite{MSD} and MM-WHS~\cite{mmwhs} datasets, respectively.

\begin{table}
	\setlength{\abovecaptionskip}{0.pt}
	\setlength{\belowcaptionskip}{-0.em}
	\centering
	\footnotesize
\begin{threeparttable}
	\begin{tabular}{cccc}
		\toprule[1.2pt]
		\multicolumn{2}{c}{\textbf{Loss Functions}} &\multirow{2}{*}{\textbf{BTCV}} &\multirow{2}{*}{\textbf{MM-WHS}}\\
        \cline{1-2}
        \bm{$L_{pred}$} &\bm{$L_{reg}$} &\\
		\hline
        \XSolidBrush &\XSolidBrush &80.53 &86.11\\
        \hline
        \CheckmarkBold &\XSolidBrush &82.96 &88.82\\
        \rowcolor{mygray}
        \CheckmarkBold &\CheckmarkBold &\textbf{83.85} &\textbf{90.54}\\
        \toprule[1.2pt]
	\end{tabular}
    \end{threeparttable}        
	\caption{Evaluation of loss functions $L_{pred}$ and $L_{reg}$. We report the average Dice Score on BTCV~\cite{btcv} and MM-WHS~\cite{mmwhs}.}
\label{table_abla_loss}
\end{table}%

\begin{table}
	\setlength{\abovecaptionskip}{0.pt}
	\setlength{\belowcaptionskip}{-0.em}
	\centering
	\footnotesize
\begin{threeparttable}
	\begin{tabular}{ccc}
		\toprule[1.2pt]
		\textbf{Number of bases} \bm{$n$} &\textbf{BTCV} &\textbf{MM-WHS}\\
        \hline
        \emph{\bm{$(x, y, z)$}}\\
        \hline
        \emph{\bm{$2{\times}2{\times}1$}} &81.56 &86.73\\
        \emph{\bm{$3{\times}3{\times}1$}} &82.77 &89.31\\
        \rowcolor{mygray}
        \emph{\bm{$4{\times}4{\times}1$}} &\textbf{83.85} &\textbf{90.54}\\
        \emph{\bm{$5{\times}5{\times}1$}} &83.60 &90.49\\
        \hline
        \emph{\bm{$3{\times}3{\times}2$}} &82.86 &89.14\\
        \emph{\bm{$4{\times}4{\times}2$}} &83.47 &90.52\\
        \toprule[1.2pt]
	\end{tabular}
    \end{threeparttable}        
	\caption{Evaluation of the value of bases $n$. We report the average Dice Score on BTCV~\cite{btcv} and MM-WHS~\cite{mmwhs}.}
\label{table_abla_number}
\end{table}%

\noindent\textbf{Generalization capacity on MRI dataset}. To verify the generalization capacity on the MRI dataset, we further evaluate the performance of VoCo on BraTS 21~\cite{brats}. As shown in Table~\ref{table_brats}, VoCo achieves 78.53\% Dice Score and outperforms existing state-of-the-art methods, demonstrating the cross-model generalization capacity of VoCo.

\noindent\textbf{Evaluation of COVID-19 classification}. We further evaluate the performance of the classification task on the CC-CCII~\cite{CC-CCII} dataset in Table~\ref{table_cc}. Since existing SSL methods did not conduct experiments on this dataset, we reproduce the related methods for comparisons. It can be seen that VoCo can also achieve superior results with 90.83\% accuracy, proving its effectiveness in the classification task.

\noindent\textbf{Overall comparisons on six downstream datasets}. The overall comparisons are shown in Fig.~\ref{fig_over}. Specifically, we compare with the existing state-of-the-art methods on six different downstream datasets. It can be seen that our VoCo outperforms them by a clear margin.

\subsection{Ablation Study}
\label{sec_ablation}

We further conduct ablation studies to evaluate the loss functions and the settings in VoCo, which are verified on the BTCV~\cite{btcv} and MM-WHS~\cite{mmwhs} datasets.

\noindent\textbf{Loss functions}. We first study the importance of the two loss functions, \emph{i.e.}, $L_{pred}$ and $L_{reg}$, as shown in Table~\ref{table_abla_loss}. It can be seen that with our proposed $L_{pred}$ loss function, the performance is significantly improved, \emph{i.e.}, 80.53\% to 82.96\% on BTCV, 86.11\% to 88.82\% on MM-WHS. These results demonstrate the effectiveness of our proposed position prediction pretext task. In addition, with the proposed regularization loss $L_{reg}$, the performance can be further improved. Thus, we can see that it is crucial to learn discriminative bases in VoCo.

\noindent\textbf{Number of bases}. We further evaluate different settings of the number of bases $n$ in VoCo. We compare with different settings of $n$ in the ablation studies, as shown in Table~\ref{table_abla_number}. It is worth noting that due to the ROI size inconsistency in the $Z$ direction, it is not practical to crop multiple bases in the $Z$ direction, since we have to resize the volume after crops, which will result in inconsistent volume scales. In addition, due to the computation limitation, it is costly to increase the values of $n$. As shown in Table~\ref{table_abla_number}, with $n=2{\times}2{\times}1$, the VoCo only achieves 81.56\% and 86.73\% Dice Score on BTCV and MM-WHS, respectively. When we increase the values of $n$ to $3{\times}3{\times}1$ and $4{\times}4{\times}1$, the performances are improved obviously. Specifically, with $n$ as $4{\times}4{\times}1$, we achieve 83.85\% and 90.54\% on BTCV and MM-WHS, respectively. However, we observe that higher $n$ ($5{\times}5{\times}1$) cannot further bring higher performance. We further verify the performance of generating base crops in the $Z$ direction. It can be seen that $3{\times}3{\times}2$ and $4{\times}4{\times}2$ cannot yield improvements. Thus, aiming to balance the performance and efficiency, we set $n$ as $4{\times}4{\times}1$ in VoCo. It can be seen that the setting of $n$ is crucial to VoCo. 

\noindent Visualization results on BTCV~\cite{btcv} are shown in Fig.~\ref{fig_btcv_vis}. It can be seen that VoCo can yield improved segmentation accuracy and completeness. More visualization results are in the supplementary materials.


\section{Conclusion and Future Directions}
\label{sec5}
In this paper, we develop a simple-yet-effective SSL framework VoCo for 3D medical image analysis. Motivated by the observation that 3D medical images contain relatively consistent contextual positions between different organs, we propose to leverage the contextual position priors to learn consistent semantic representations in pre-training. Specifically, we crop volumes from different positions in an input volume and represent them as a group of bases to represent features in different directions. Then, we predict the contextual position of a randomly cropped volume by contrasting its similarity to different bases. In this way, VoCo effectively encodes the contextual position priors into model representations, enabling us to effectively improve the performance of downstream tasks that require high-level semantics. Extensive experiments demonstrate that VoCo achieves superior performance.


\noindent We will further consider several ways of extension: (1) Scale up the pre-training dataset to evaluate the upper performance of VoCo. (2) Experiments on more downstream datasets. (3) Evaluate the label-efficient performance of VoCo (\emph{e.g.}, semi-supervised learning). (4) Explore the capacity of VoCo in mining inter-volume relationships.

\section*{Acknowledgments}
\noindent This work was supported by Hong Kong Innovation and Technology Fund (Project No.~ITS/028/21FP and No. MHP/002/22), and Research Grants Council of the Hong Kong Special Administrative Region, China (Project No. T45-401/22-N).

\section{Appendix}

In the supplementary materials, we first introduce the pre-training and downstream datasets we use in our experiments. Then, we present the implementation details of VoCo, including the settings of pre-processing, pre-training, and finetuning. Finally, \textbf{additional experiments} are presented, including ablation studies and experiments on \textbf{2D medical dataset}~\cite{chestx-ray}.

\renewcommand{\thesubsection}{A}
\subsection{Datasets}

\textbf{Pre-training and downstream datasets.} The details of pre-training and downstream datasets are shown in Table~\ref{table_dataset}. Specifically, we use BTCV~\cite{btcv} and TCIA Covid19~\cite{tcia} totally about 0.8k CT scans for BTCV~\cite{btcv} downstream task, which aims to conduct fair comparison with previous works~\cite{MAE3D, GLMAE}. And we further combine LUNA~\cite{luna} to scale the size of pre-training datasets to 1.6k for the other four downstream tasks.

\textbf{BTCV dataset.} BTCV~\cite{btcv} dataset contains one background class and thirteen organ classes, \emph{i.e.}, spleen, right kidney, left kidney, gallbladder, esophagus, liver, stomach, aorta, inferior vena cava, portal and splenic veins, pancreas, left and right adrenal glands. Following the previous works~\cite{MAE3D,GLMAE,Modelgen,swin}, we split BTCV~\cite{btcv} dataset into 24 scans for training and 6 scans for validation. It is worth noting that the BTCV~\cite{btcv} dataset is used in pre-training. 

\begin{table}[!h]
	\setlength{\abovecaptionskip}{0.pt}
	\setlength{\belowcaptionskip}{-0.em}
	\centering
	\footnotesize
\begin{threeparttable}
	\begin{tabular}{ccccc}
		\toprule[1.2pt]
		\textbf{Dataset} &\textbf{Modality} &\textbf{Task} &\textbf{Train} &\textbf{Valid.}\\
		\hline
        \textbf{\emph{Pre-training}}\\
        BTCV~\cite{btcv} &CT &pre-train &24 &6\\
        TCIA Covid19~\cite{tcia} &CT &pre-train &722 &49\\
        LUNA~\cite{luna} &CT &pre-train &843 &45\\
        \hline
        \textbf{\emph{Downstream}}\\
        BTCV~\cite{btcv} &CT &segmentation &24 &6\\
        LiTs~\cite{lits} &CT &segmentation &100 &31\\
        MSD Spleen~\cite{MSD} &CT &segmentation &32 &9\\
        MM-WHS~\cite{mmwhs} &CT &segmentation &14 &6\\
        BraTs~\cite{brats} &MRI &segmentation &387 &97\\
        CC-CCII~\cite{CC-CCII} &CT &classification &2514 &1664\\
        \toprule[1.2pt]
	\end{tabular}
    \end{threeparttable}        
	\caption{The details of pre-training and downstream datasets.}
\label{table_dataset}
\end{table}%

\begin{table}
	\setlength{\abovecaptionskip}{0.pt}
	\setlength{\belowcaptionskip}{-0.em}
	\centering
	\footnotesize
\begin{threeparttable}
	\begin{tabular}{cc}
		\toprule[1.2pt]
        \textbf{\emph{Pre-process settings}}\\
        \hline
        Spacing &[1.5, 1.5, 1.5]\\
        Norm [$a_{min}, a_{max}$] &[-175.0, 250.0]\\
        Norm [$b_{min}, b_{max}$] &[0.0, 1.0]\\
        Roi-Size &$64{\times}64{\times}64$\\
        Augmentation &Random rotate and flip\\
        \toprule[1.2pt]
        \textbf{\emph{Pre-training settings}}\\
        \hline
        Pre-training steps &100k\\
        Optimizer &AdamW\\
        Optimization LR &1e-3\\
        LR schedule &warmup cosine\\
        Warmup steps &100\\
        Momentum &0.9\\
        Regularization weight &1e-2\\
        Batch size &4\\
        Sw batch size &4\\
        VoCo Resize &$384{\times}384{\times}64$\\
        Resize after crop &$64{\times}64{\times}64$\\
        VoCo $n$ &$4{\times}4$\\
        VoCo $\lambda$ &1.0\\
        \toprule[1.2pt]
        \textbf{\emph{Finetuning settings}}\\
        \hline
        Optimizer &AdamW\\
        Optimization LR &3e-4\\
        LR schedule &warmup cosine\\
        Warmup steps &100\\
        Momentum &0.9\\
        Regularization weight &1e-5\\
        Batch size &1\\
        Sw batch size &4\\
        Inference &sliding window\\
        ROI size &$96{\times}96{\times}96$\\
        \toprule[1.2pt]
	\end{tabular}
    \end{threeparttable}        
	\caption{Pre-process and training settings in the experiments.}
\label{table_settings}
\end{table}%

\textbf{LiTs dataset.} LiTs~\cite{lits} dataset releases 131 abdominal CT Volumes and associated annotations for training and validation. There are two types of labels in LiTs~\cite{lits}: the liver and tumor. Following previous works~\cite{PCRLv1, PCRLv2, dive}, in this paper, we only utilize the ground truth masks of the liver to evaluate the effectiveness of various SSL algorithms. 

\textbf{MSD Spleen dataset.} MSD Spleen dataset is the $9_{th}$ challenge in MSD~\cite{MSD}, which is developed for spleen segmentation. Specifically, aiming to conduct fair comparisons with previous state-of-the-art methods~\cite{geo,MAE3D,GLMAE}, we use 32 scans for training and 9 scans for validation, as shown in Table~\ref{table_dataset}.

\textbf{MM-WHS dataset.} MM-WHS~\cite{mmwhs} dataset is also unseen in the pre-training, which contains 7 classes including Left
Ventricle, whole aorta, Right Ventricle, Left Atrium, myocardium of Left Ventricle, Right Atrium, and Pulmonary Artery. The data splits are also shown in Table~\ref{table_dataset}.

\textbf{BraTs dataset.} BraTs~\cite{brats} dataset is an MRI dataset, which has been known as a series of challenges in brain tumor segmentation. In this paper, we evaluate the ability of model generalization on the BraTs~\cite{brats} dataset, since we pre-train the model with only CT datasets. Specifically, we perform experiments on the released 387 scans
of BraTS 2021 and evaluate the accuracy on the remained 97 scans. There are three classes in BraTS: whole tumor (WT), tumor core (TC), and enhancing tumor (ET). 

\begin{figure*}
	\centering
	\includegraphics[width=1\linewidth]{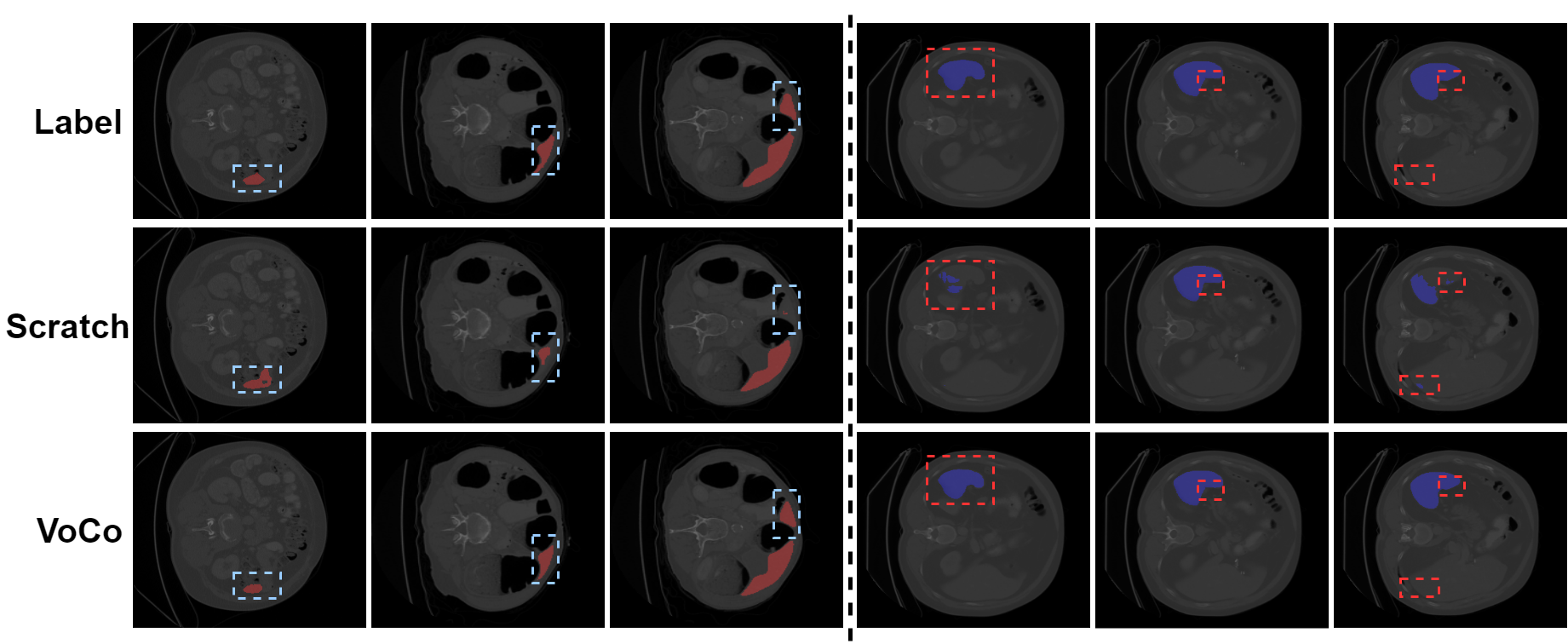}
	\caption{Qualitative visualization of segmentation results for the LiTS~\cite{lits} and MSD Spleen~\cite{MSD} datasets. Scratch represents the results of `from scratch'. The obvious differences are highlighted by blue and red dashed boxes, respectively.}
	\label{fig_lits_vis}
\end{figure*}

\begin{figure*}
	\centering
	\includegraphics[width=1\linewidth]{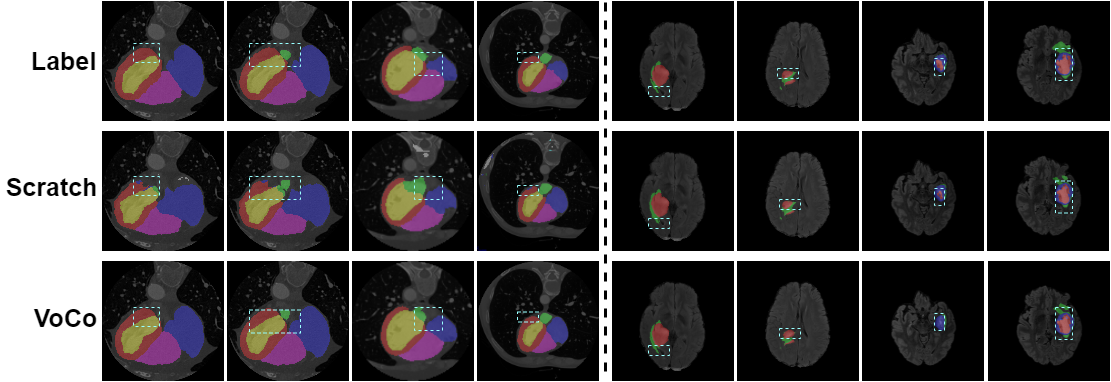}
	\caption{Qualitative visualization of segmentation results for the MM-WHS~\cite{mmwhs} and BraTs~\cite{brats} datasets. Scratch represents the results of `from scratch'. The obvious differences are highlighted by blue dashed boxes, respectively.}
	\label{fig_whs_bra_vis}
\end{figure*}

\textbf{CC-CCII dataset.} CC-CCII~\cite{CC-CCII} dataset is designed for COVID-19 detection, which can be seen as a classification task. CC-CCII~\cite{CC-CCII} dataset contains 2516 scans for training and 1644 scans for validation, which includes three classes, \emph{i.e.}, novel coronavirus pneumonia (NCP), common pneumonia (CP), and normal controls (Normal).

\renewcommand{\thesubsection}{B}
\subsection{Implementation Details}

Aiming to conduct fair comparisons with previous methods~\cite{PCRLv2,dive,geo,GLMAE,MAE}, we adopt comparatively consistent settings in the experiments. The details of pre-process and training settings are shown in Table~\ref{table_settings}. Our implementation is mainly based on the open-source platform Monai~\footnote{\href{https://monai.io/}{https://monai.io/}} and Pytorch~\cite{pytorch}. We use one NVIDIA A100 GPU for all the experiments.

\textbf{Fine-tuning on downstream datasets}. The fine-tuning settings are almost consistent with the pre-training settings, except for the number of training epochs. Specifically, the training epochs are set to 3000, 1000, 1000, 1000, 500, and 100 for BTCV~\cite{btcv}, LiTs~\cite{lits}, MSD Spleen~\cite{MSD}, MM-WHS~\cite{mmwhs}, BraTs~\cite{brats}, and CC-CCII~\cite{CC-CCII}, respectively.

\renewcommand{\thesubsection}{C}
\subsection{Experiments}

We provide some experiments that are not presented in the main paper due to the limitation of pages, including ablation studies, 2D medical image analysis, and others.

\subsubsection{Ablation Studies}
We further evaluate the settings of the balance parameter $\lambda$ for the loss functions, as shown in Table.~\ref{table_lambda}. We also report the Dice Score on the BTCV~\cite{btcv} and MM-WHS~\cite{mmwhs} datasets for evaluation. We set $\lambda$ as 0.5, 1.0, and 1.5 for ablation studies. As shown in Table.~\ref{table_lambda}, we find that the settings of $\lambda$ do not matter a lot. Thus, in VoCo, we consider the importance of loss functions equal and set $\lambda$ as 1.

\subsubsection{2D Medical Image Analysis}

\begin{table}
	\setlength{\abovecaptionskip}{0.pt}
	\setlength{\belowcaptionskip}{-0.em}
	\centering
	\footnotesize
\begin{threeparttable}
	\begin{tabular}{c|cc}
		\toprule[1.2pt]
        \bm{$\lambda$} &\textbf{BTCV} &\textbf{MM-WHS}\\
        \hline
        0.5 &83.52 &90.16\\
        1.0 &\textbf{83.85} &\textbf{90.54}\\
        1.5 &83.80 &90.48\\ 
        \toprule[1.2pt]
	\end{tabular}
    \end{threeparttable}        
	\caption{Ablation studies of $\lambda$ on BTCV~\cite{btcv} and MM-WHS~\cite{mmwhs}.}
\label{table_lambda}
\end{table}%

\begin{table}
	\setlength{\abovecaptionskip}{0.pt}
	\setlength{\belowcaptionskip}{-0.em}
	\centering
	\footnotesize
\begin{threeparttable}
	\begin{tabular}{ccc}
		\toprule[1.2pt]
        \textbf{Methods} &\textbf{NIH ChestX-ray}\\
        \hline
        From scratch &75.4\\
        \hline
        MG~\cite{Modelgen} &77.3\\
        TransVW~\cite{TransVW} &77.6\\
        C2L~\cite{C2L} &79.0\\
        SimSiam~\cite{simsiam} &79.4\\
        PCRLv1~\cite{PCRLv1} &79.9\\
        PCRLv2~\cite{PCRLv2} &81.5\\
        \rowcolor{mygray}
        \textbf{VoCo} &\textbf{82.02}\\
        \toprule[1.2pt]
	\end{tabular}
    \end{threeparttable}        
	\caption{Experimental results on the NIH ChestX-ray~\cite{chestx-ray} dataset. The results are drawn from~\cite{PCRLv2}.}
\label{table_chestx}
\end{table}%

\begin{table}
	\setlength{\abovecaptionskip}{0.pt}
	\setlength{\belowcaptionskip}{-0.em}
	\centering
	\footnotesize
\begin{threeparttable}
	\begin{tabular}{cc}
		\toprule[1.2pt]
        \textbf{Organs} &\textbf{Dice Scores(\%)}\\
        \hline
        Left Ventricle &91.32\\
        Whole aorta &91.30\\
        Right Ventricle &94.64\\
        Left Atrium &86.89\\
        Myocardium of Left Ventricle &89.16\\
        Right Atrium &96.35\\
        Pulmonary Artery &84.13\\
        \hline
        \textbf{Average} &\textbf{90.54}\\
        \toprule[1.2pt]
	\end{tabular}
    \end{threeparttable}        
	\caption{Dice Scores of 7 organs on MM-WHS~\cite{mmwhs}.}
\label{table_whs}
\end{table}%

In the main paper, we evaluate the effectiveness of VoCo on 3D medical images. To further verify its performance on 2D medical images, we also conduct experiments on the NIH ChestX-ray~\cite{chestx-ray} datasets. We follow the consistent settings of previous works~\cite{PCRLv1,PCRLv2}, \emph{i.e.}, pre-train on NIH ChestX-ray and fine-tune on NIH ChestX-ray. Specifically, for fair comparisons with~\cite{PCRLv1,PCRLv2}, 60\% of data are used for pre-training and the remaining is used for finetuning. 3D-UNet~\cite{UNET} is used for experiments. As shown in Table~\ref{table_chestx}, VoCo can also achieve competitive results on the 2D medical dataset. We conclude that although the 2D images contain less information than 3D scans, the position priors still exist, which benefits the training of VoCo.

\subsubsection{Dice Scores of MM-WHS dataset}

The Dice Scores of 7 organs on the MM-WHS~\cite{mmwhs} dataset are shown in Table~\ref{table_whs}.

\subsubsection{Validation results on the leaderboard}

We have verified the \emph{\textbf{BTCV}} test results and further evaluated the test sets of \emph{\textbf{Flare23}} and \emph{\textbf{Amos22}} in the public leaderboard. \textbf{Note that} aiming to verify the pure effectiveness, we did not use model ensembling, extra data, or other tricks. We compare with the strong baseline SwinUNETR\cite{swinunetr} (since with the same network and settings) in Table~\ref{table_board}. 

The MSD leaderboard has not been updated for a long time. Due to the rebuttal emergency, we provide the results of the offline validation set instead. We strictly follow the settings of SwinUNETR\cite{swinunetr} and the results are shown in Table~\ref{table_msd2}. 

\begin{table}[!h]
	\setlength{\abovecaptionskip}{0.pt}
	\setlength{\belowcaptionskip}{-0.em}
	\centering
	\footnotesize
\begin{threeparttable}
	\begin{tabular}{c|ccc}
		\toprule[1.2pt]
		\textbf{Method} &\textbf{BTCV} &\textbf{Flare23} &\textbf{Amos22} (DSC/NSD)\\
        \hline
        SwinUNETR\cite{swinunetr} &$\dagger$84.72 &87.84 &$\dagger$88.00/76.15\\
        \rowcolor{mygray} 
        \textbf{VoCo} &\textbf{86.44} &\textbf{90.07} &\textbf{89.06/78.86}\\
        \toprule[1.2pt]
	\end{tabular}
    \end{threeparttable}        
	\caption{\textbf{Online test results}. $\dagger$: drawn from previous papers.}
\label{table_board}
\end{table}%

\begin{table}[!h]
	\setlength{\abovecaptionskip}{0.pt}
	\setlength{\belowcaptionskip}{-0.em}
	\centering
	\footnotesize
\begin{threeparttable}
	\begin{tabular}{c|ccccc}
		\toprule[1.2pt]
		\textbf{Method} &\textbf{Task1} &\textbf{Task2} &\textbf{Task3} &\textbf{Task4} &\textbf{Task5}\\
        \hline
        SwinUNETR$\dagger$[47] &75.13 &95.89 &81.72 &91.98 &80.23 \\
        \rowcolor{mygray} 
        \textbf{VoCo} &\textbf{76.26} &\textbf{96.93} &\textbf{84.98} &\textbf{92.09} &\textbf{82.16}\\
	\end{tabular}
    \end{threeparttable}
\label{table_msd1}
\end{table}%
\vspace{-.3in}
\begin{table}[!h]
	\setlength{\abovecaptionskip}{0.pt}
	\setlength{\belowcaptionskip}{-0.em}
	\centering
	\footnotesize
\begin{threeparttable}
	\begin{tabular}{c|ccccc}
		\textbf{Method} &\textbf{Task6} &\textbf{Task7} &\textbf{Task8} &\textbf{Task9} &\textbf{Task10}\\
        \hline
        SwinUNETR$\dagger$\cite{swinunetr} &63.46 &64.32 &70.54 &94.63 &44.57\\
        \rowcolor{mygray} 
        \textbf{VoCo} &\textbf{67.74} &\textbf{67.85} &\textbf{70.92} &\textbf{96.34} &\textbf{45.17}\\
        \toprule[1.2pt]
	\end{tabular}
    \end{threeparttable}        
	\caption{MSD Decathlon. More results will be in the revision.}
\label{table_msd2}
\end{table}%
\vspace{-.15in}

\subsubsection{More Visualization Results}

Visualization results on LiTs~\cite{lits}, MSD Spleen~\cite{MSD}, MM-WHS~\cite{mmwhs}, and BraTs~\cite{brats} are shown in Fig.~\ref{fig_lits_vis} and Fig.~\ref{fig_whs_bra_vis}.

{\small
\bibliographystyle{ieee_fullname}
\bibliography{egbib}
}

\end{document}